\documentclass[12pt,a4paper]{article}
\usepackage[margin=1in]{geometry}
\usepackage{graphicx}
\usepackage{amsmath, amssymb}
\usepackage{newtxtext}
\usepackage{newtxmath}
\usepackage{siunitx}
\usepackage{authblk}
\usepackage{xcolor} 
\definecolor{redstrikin}{rgb}{1,0,0}
\usepackage[font=small,labelfont=bf]{caption}
\usepackage[superscript,biblabel]{cite}
\usepackage{float}
\usepackage[colorlinks=true,
    linkcolor=red,
    citecolor=red,
    urlcolor=blue,
    filecolor=blue
]{hyperref}
\usepackage{subcaption}
\usepackage{setspace}
\usepackage{titlesec}
\newcommand{\rv}[1]{\textcolor{black}{#1}}

\titleformat{\section}
  {\sffamily\bfseries\large}
  {\thesection.}{1em}{}

\titleformat{\subsection}
  {\sffamily\bfseries\normalsize}
  {\thesubsection.}{1em}{}

\title{\textbf{Interfacial standing wave-patterns disentangle dilatational and shear surface viscous effects}}

\author[1]{Debashis Panda}
\author[1,2]{Abdullah M. Abdal}
\author[1]{Mosayeb Shams}
\author[1]{Lyes Kahouadji}
\author[3]{Jalel Chergui}
\author[4]{Seungwon Shin}
\author[3,5]{Damir Juric}
\author[1,*]{Omar K. Matar}

\affil[1]{Matar Fluids Group, Department of Chemical Engineering, Imperial College London, London SW7 2AZ, United Kingdom}
\affil[2]{Department of Environmental and Sustainability Engineering, College of Engineering and Energy, Abdullah Al Salem University, Kuwait}
\affil[3]{Université Paris Saclay, CNRS, Laboratoire Interdisciplinaire des Sciences du Numérique (LISN), Orsay, France}
\affil[4]{Department of Mechanical and System Design Engineering, Hongik University, Seoul, Republic of Korea}
\affil[5]{Department of Applied Mathematics and Theoretical Physics, University of Cambridge, Cambridge CB3 0WA, UK}
\affil[*]{Corresponding author: omar.matar@imperial.ac.uk}

\date{} 
\begin{document}
\maketitle
\begin{abstract}
    \rv{Dilatational and shear surface viscosities are highly correlated parameters, making their individual contributions difficult to disentangle in Stokes flow, linearised flow models, or two-dimensional flows.}
    We therefore investigate the \rv{three-dimensional interfacial standing waves as a means to decouple the influence of dilatational and shear surface viscosities.} Two dimensionless controlling parameters are introduced: $Bq$, the total Boussinesq number, which quantifies the the relative importance of surface viscous stresses compared with bulk viscous stresses, and $\tan \chi$, which quantifies the ratio of \rv{surface dilatational viscosity to surface shear viscosity.} 
    The growth rates and threshold accelerations are independent of $\chi$, consistent with previous theoretical predictions. Nonlinear analyses of square and hexagonal patterns reveal that Fourier decomposition of wave-patterns can effectively decouple the intricate dynamics into axial modes, where the waves are \rv{weakly dependent} on $\chi$, and oblique modes, where additional damping occurs in the shear surface viscous dominant interface. 
    These results demonstrate that Faraday wave-patterns provide a route for identifying and quantifying the distinct roles of dilatational and shear surface viscosities. 
\end{abstract}

\section*{Introduction}
Surface-active agents, such as proteins, lipids, polymers, or surfactants, adsorb at fluid--fluid interfaces and alter their mechanical response by introducing in-plane friction \cite{fuller2012complex, miller2010rheology, choi2011active}. The resulting molecular rearrangements give rise to surface rheology \cite{jaensson2021computational}, which couples interfacial mechanics \rv{with} bulk flow dynamics. Such complex interfaces underpin diverse phenomena, from aerosol generation during bubble collapse \cite{ji2023secondary, tammaro2021flowering}, which can influence climate, to the deformation of biological membranes that governs solute transport\cite{erni2011deformation}. Despite the broad relevance of surface rheology, systematic studies on its effects remain limited. Most of the literature is restricted either to linear flow regimes \cite{lopez1998direct, flumerfelt1980effects} or to two-dimensional flows \cite{singh2020deformation, ubal2005influence, wee2022absence}, which neglects the key three-dimensional complexities of surface rheological dynamics in real-world applications. Here, we present a novel investigation of three-dimensional dynamics of two-dimensional standing wave-patterns to isolate and quantify surface-rheological effects at complex interfaces.  

Standing waves that arise at a fluid interface subjected to vertical vibration, commonly referred to as Faraday waves, constitute one of the simplest and most accessible systems for investigating out-of-equilibrium dynamics. In laterally unbounded configurations, these interfacial waves exhibit directional isotropy and interact through spatial superposition, generating a rich variety of ordered patterns. Depending on the underlying symmetries, the system may be organised into simple periodic structures, such as squares and hexagons \cite{perinet2009numerical} or more intricate arrangements, including double hexagons \cite{edwards1993parametrically} and superlattices \cite{kudrolli1998superlattice}. In the presence of surface-active agents, surface waves not only experience enhaced \rv{damping} \cite{henderson1998effects}, but can also significantly modify the resulting wave patterns \cite{ merkt2004persistent, panda2025marangoni}.  

The surface rheology of a Newtonian interface is commonly described by the Boussinesq–Scriven constitutive relation \cite{scriven1960dynamics}, which expresses the interfacial viscous traction in terms of the dilatational and shear surface viscosities. Understanding when dilatational versus shear surface-viscous stresses dominate is crucial for elucidating complex interfacial rheology. For instance, predominantly dilatational surface-viscous stresses can lead to a highly compressed cavity collapse in bursting bubbles, promoting secondary bubble entrainment \cite{ji2023secondary}, whereas the dominance of shear surface viscosity over dilatational surface viscosity can stabilise nonlinear flows \cite{singh2020deformation, Gounley_Boedec_Jaeger_Leonetti_2016}. Similarly, a perturbed drop has been found to elongate and destabilise into daughter droplets when the dilatational surface viscosity is sufficiently large. Consistent with this variability, experimental measurements report that the ratio of dilatational to shear surface viscosities can span up to five orders of magnitude 
\cite{erni2011deformation, erni2011emulsion, kragel1996surface}. This implies that reliable quantification of both surface viscosities is essential for characterising complex interfacial rheology.   

Numerous experimental strategies have been developed to measure surface viscosities, but they face significant limitations, primarily due to coupling between surface viscosity and surface tension gradients \cite{miller2010rheology, choi2011active, fuller2012complex}. 
Theoretical studies have provided partial insights, yet are often hindered by inherent simplifying assumptions, particularly those associated with linearised flow models, Stokes flow, or two-dimensional flows. The origin of these theoretical limitations can be seen by examining the surface-excess traction that appears in the interfacial force balance. In the Boussinesq--Scriven framework, the surface-excess traction ($\mathbf n \cdot \mathbf P$, where $\mathbf n$ and $\mathbf P$ are the normal and surface-excess tensor to the interface) due to varying surface tension ($\sigma$) and rheology ($\mu_d$ and $\mu_s$ for dilatational and shear surface viscosities, respectively) for a three-dimensional interface is given by, 
$\big[ \mathbf{n} \cdot \mathbf{P} \big]_{\mathrm{3D}}
=
\big[ \mathbf{n} \cdot \mathbf{P} \big]_{\mathrm{2D}}
+ \mu_s\, f(\mathbf n,\mathbf u),$
where 
\rv{$\mathbf u$ is the interfacial velocity, $\nabla_s = (\mathbf I-\mathbf{nn})\nabla$ is the surface gradient operator, $\mathbf I$ is the identity tensor, and $f(\mathbf n, \mathbf u) = - (\mathbf{n} \times \nabla_s \mathbf{n} \times \mathbf{n})
: \nabla_s \mathbf{n}\, \mathbf{u}
+ \mathbf{n} \times \nabla_s
\big( \mathbf{n} \cdot \nabla_s \times \mathbf{u} \big)
+ 2(\mathbf{n} \times \nabla_s \mathbf{n} \times \mathbf{n})
\cdot \nabla_s (\mathbf{u} \cdot \mathbf{n})
- 2\,\mathbf{n}\,
(\mathbf{n} \times \nabla_s \mathbf{n} \times \mathbf{n})
: \nabla_s \mathbf{u}$ is a highly nonlinear term that depends on the interfacial curvature, normal, and velocity.}
The nonlinear function $f(\mathbf n, \mathbf u)$ is significant if and only if $\mu_s \neq 0$. 
In contrast, in linearised and two-dimensional settings, $f(\mathbf n, \mathbf u) \simeq 0$, and thus the expression simplifies such that the surface viscosities enter only through a combination of dilatational and shear surface viscosities, 
$[\mathbf n \cdot \mathbf P]{_{2\rm D}} = \mathbf n\big(\sigma + (\mu_d + \mu_s)(\nabla_s \cdot \mathbf u)\big)\kappa + 
\nabla_s \sigma + (\mu_d + \mu_s)\nabla_s (\nabla_s \cdot \mathbf u)$ \cite{brenner2013interfacial}.
Although dilatational surface viscosity \rv{influences} surface compressibility and shear surface viscosity affects the surface velocity gradients at the interface, their contributions in such flows are highly correlated and therefore difficult to disentangle. Moreover, it is challenging to determine the dominance of surface viscosity effects when they appear as a correlated parameter. Motivated by this, in this study, we investigate the significance of the three-dimensional surface rheological effects in the dominant regimes of dilatational and shear surface viscosities in the context of standing wave-patterns.

In this study, we show that the distinction between the two surface viscosities can be tracked via patterns formed on an interface under vertical vibration. The shear surface viscosity provides additional damping by suppressing the surface velocity gradients that occur due to the flow from the trough to the crest. The damping due to the shear surface viscosity is magnified in the case of a six-fold symmetric domain, and a pattern transition from a six-fold symmetric hexagon to a two-fold symmetric broken hexagon is observed as we increase (decrease) the shear (dilatational) surface viscosity. The proof-of-concept presented in this study can be utilised as a novel strategy to decouple the shear and dilatational surface viscous effects on a complex interface.
\section*{Results and discussion}
We consider a three-dimensional cuboidal computational domain that is laterally periodic and bounded by closed top and bottom boundaries, with a fixed height $H = 5~\rm mm$. As shown in figure~\ref{fig_1}(a), three different lateral side configurations are chosen to satisfy two-fold, or quasi-2D ($\lambda_c \times \lambda_c/4$), four-fold square ($\lambda_c \times \lambda_c$), and six-fold hexagon ($2\lambda_c \times 2\lambda_c/\sqrt 3$) symmetries, where $\lambda_c$ is the critical wavelength of the surface waves when the interface is clean. In all configurations, the domain is filled to a depth $h=1~\rm{mm}$ in all configurations with a water--glycerin mixture, initially flat and at rest, with air above. In the inertial frame of reference, vibration appears as a parametric modulation of the volumetric forces at an angular frequency $\Omega$ and an amplitude $A$. The nondimensional governing equations involve the Reynolds, $Re$, Weber, $We$, and Froude, $Fr$, numbers, which quantify the relative importance of inertial forces compared with viscous, capillary, and gravitational forces, respectively \rv{(see `Methods' in the Supplementary Information)}. The surface-rheology control parameters of interest are the shear and dilatational Boussinesq numbers, $Bq_s$ and $Bq_d$, which quantify the relative importance of surface shear and dilatational viscous stresses compared with bulk viscous stresses. Due to the entangled nature of $\mu_s$ and $\mu_d$ in linearised/two-dimensional settings, as shown in figure~\ref{fig_1}(b), we parameterise the surface rheology control space in terms of the total Boussinesq number $Bq=Bq_s+Bq_d$ and the mixing angle $\tan\chi=Bq_d/Bq_s$, to quantify the relative strength of both surface viscosities. Details of the governing equations, non-dimensionalisation, fluid properties, and numerical implementation  are provided in the Supplemental Information.

\subsection*{Linear regime and threshold acceleration}
\begin{figure}
    \centering
    \includegraphics[width=1\linewidth]{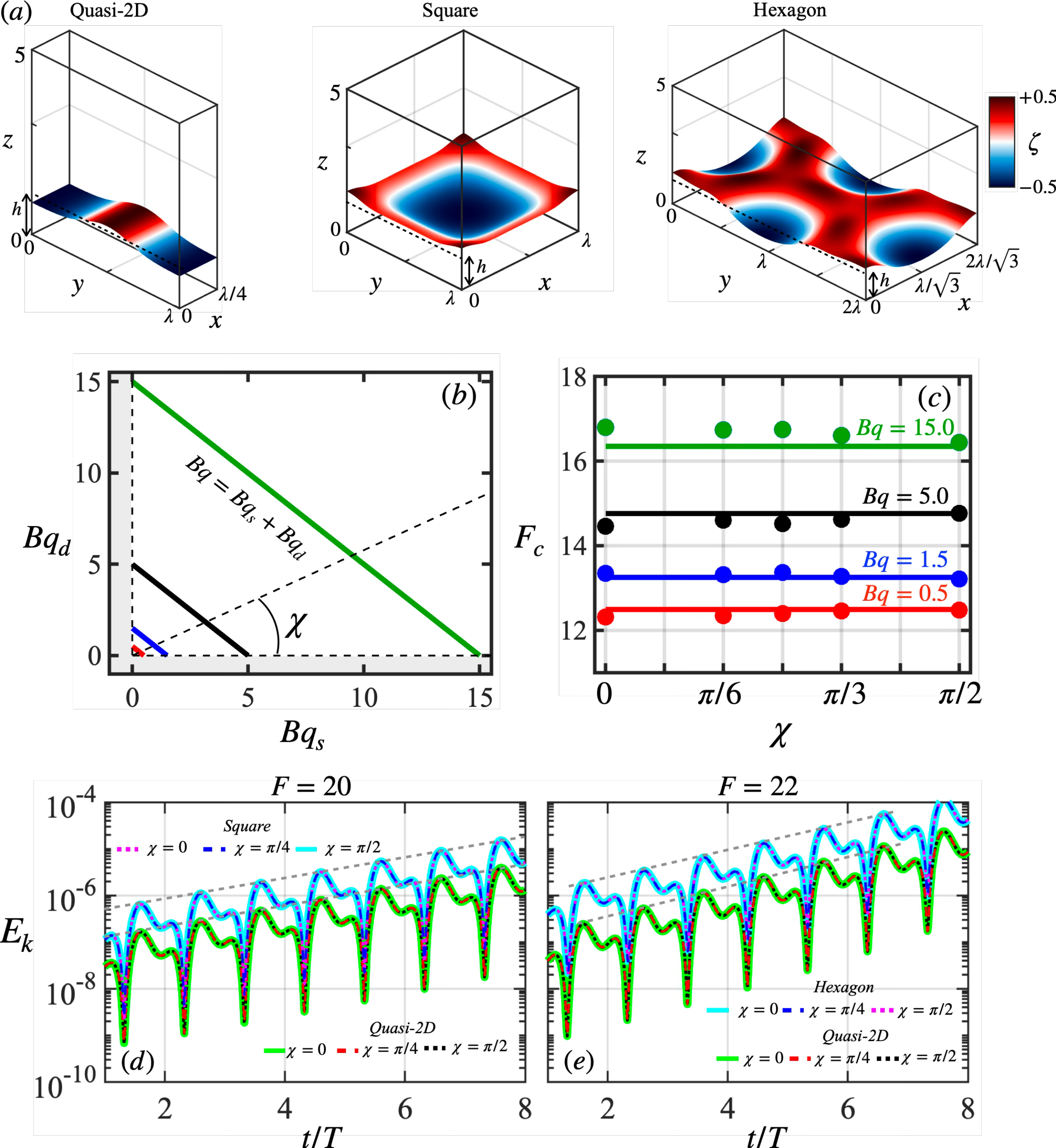}
    \caption{(a) Physical space representation of quasi-2D, square, and hexagon patterns in Faraday waves. A three-dimensional interface coloured by the interfacial height \rv{displacement $\zeta$ from the mean position, $h$} is shown in the computational domain. The regime map of the two controlling parameters, $Bq$ and $\chi$, is shown in (b), and the threshold acceleration amplitude to destabilise the interface is shown in (c) for varying $\chi$ and $Bq = 0.5, 1.5, 5, 15$. \rv{Here, the lines correspond to the threshold acceleration report in the literature for two-dimensional Faraday waves \cite{ubal2005SV} and the circles correspond to our simulation for varying $\chi$.}  
    The growth of kinetic energy is shown in (d) and (e) for $F = 20$ and $22$, respectively, for the squares and hexagons. The growth rates are compared with their respective cases in Quasi-2D computational domain. Here, $T=2\pi/\Omega$ is time period of external vibration.}
    \label{fig_1}
\end{figure}
\label{sec:results}
We begin by examining the linear growth regime of the surface waves. Using the standard procedure for determining the threshold acceleration amplitude, $F = A/g$, (with $g$ the gravitational acceleration) \cite{perinet2009numerical, panda2025marangoni}, we validated our numerically obtained threshold accelerations against the critical accelerations reported in the literature \cite{ubal2005SV}. For a range of mixing angles, $\chi$, we computed the threshold acceleration at $Bq = 0.5, 1.5, 5,$ and $15$, as shown in figure~\ref{fig_1}(c), where the solid lines correspond to the threshold acceleration obtained in the literature \cite{ubal2005SV}. The threshold acceleration is therefore independent of $\chi$ and depends only on the total Boussinesq number $Bq$, confirming that the shear and dilatational surface viscosities are not separately identifiable from the threshold acceleration criterion. 

To further verify this, we set $Bq=5$ in our investigation. For $Bq=5$, the critical acceleration amplitude required to destabilise the interface is $F_c=14.76$ \cite{ubal2005SV}. We therefore consider $F=20$ and $22$ and examine the growth phase across different lateral configurations. We assess the role of $\chi$ using $\chi = 0$ (purely shear), $\chi = \pi/2$ (purely \rv{dilatational}), and $\chi = \pi/4$ (equal shear and dilatational contributions).      
For $F > F_c$, the surface waves undergo positive growth. The temporal evolution of the kinetic energy, $E_k$, is shown in figures~\ref{fig_1}(d) and (e). The linear slope in the kinetic energy confirms that the growth phase is not influenced by the nonlinear surface-viscous effects. Moreover, for fixed $Bq$, varying $\chi$ does not affect the linear growth phase cross the three lateral configurations considered. This confirms that the initial growth of surface waves is independent of $\chi$ and that surface-viscous effects are difficult to assess during the growth phase because they appear as an additional effective surface-tension-like contribution. Moreover, the similar growth phase observed for square and hexagonal patterns suggests that the growth is two-dimensional in nature. Since the linear growth phase is independent of three-dimensional effects, linear analysis is viable for obtaining the threshold acceleration. 

\subsection*{Square standing wave-patterns} In dynamical systems such as parametric surface waves, the linear growth phase saturates as nonlinear surface wave interactions become important. This nonlinear saturation reorganises the standing waves into ordered patterns, such as squares and hexagons. The next step in our investigation is to understand how surface rheology affects parametric standing waves with square and hexagonal symmetries. 

We begin by discussing the square standing wave patterns, as shown in the middle panel of figure~\ref{fig_1}(a). The interface is \rv{coloured by the vertical displacement of the interface from the initial height $h$, given by $\zeta(\mathbf x, t) = z_f(x_f, y_f, t)-h$, where $(x_f, y_f, z_f)$ denotes the interface location in the three-dimensional space.} The snapshot is extracted at $t = 35T$, where a trough is observed at the centre of the domain 
The crests occur at the edges and corners of the periodic domain.
The spatial Fourier decomposition of the square standing wave-pattern is carried out by \rv{setting} $\zeta(\mathbf x, t) = \sum_{m,n} e^{i\mathbf k_c \cdot \mathbf x} ~\hat \zeta_{mn} $, where, $\mathbf k_c$ is the fundamental wave vector and $\hat \zeta_{m,n}$ is the $m^{\rm th}$ in $x-$ and $n^{\rm th}$ in $y-$ direction Fourier mode, given by, $\hat \zeta_{mn} = mk_c ~\mathbf e_x + nk_c ~ \mathbf e_y$, \rv{where $\mathbf e_x$ and $\mathbf e_y$ are the unit vectors in $x$ and $y$ directions, respectively.}  

Fourier modes along the $x-$ ($n = 0$) and $y-$ ($m = 0$) axes are the axial modes, which decompose the two-dimensional surface wave-pattern (or three-dimensional flow) into one-dimensional waves (or two-dimensional flow). In square symmetry (two-fold symmetry), the non-negligible axial modes are $\hat \zeta_{01}$ and $\hat \zeta_{10}$ of wave number $k_c$. The superposition of these waves leads to a non-negligible oblique Fourier mode, $\hat \zeta_{11} = \hat \zeta_{10} + \hat \zeta_{01}$ of wave number $\sqrt 2 k_c$ \cite{perinet2009numerical}. 
\rv{Since, this is a oblique mode where the wave vector is directed at an angle of $\pi/4$ in Fourier space, the three-dimensional surface rheological effects are expected.} Thus, $\hat \zeta_{11}$ is a key indicator for assessing the three-dimensional effects of surface viscosity. 

\begin{figure*}[!ht]
    \centering
    \includegraphics[width=1.0\linewidth]{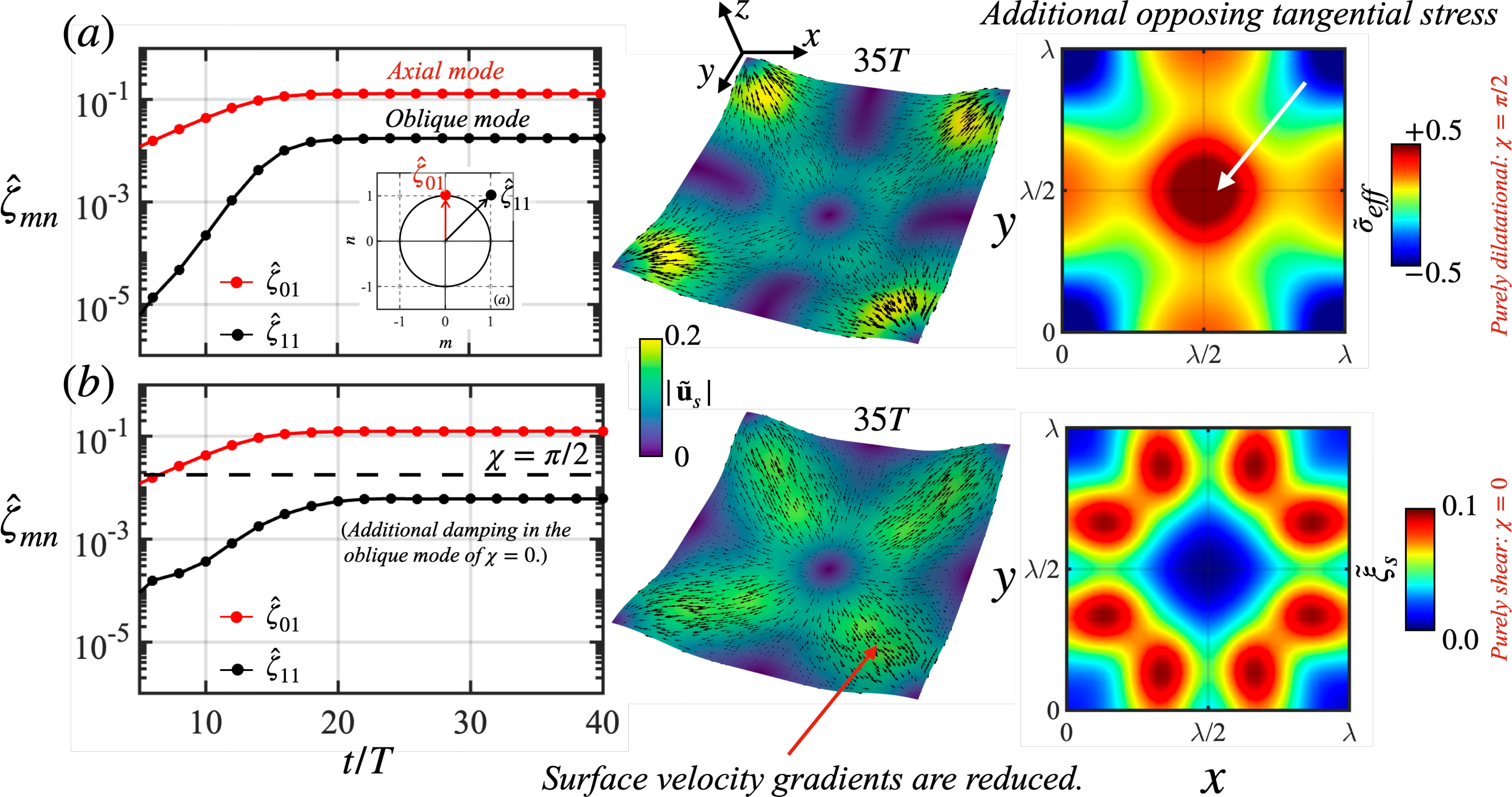}
    \caption{Temporal evolution of the axial (red) and oblique (black) modes (see the inset in (a) for the description of the Fourier space) in square wave-pattern is shown for $\chi = \pi/2$ in panel (a) and $\chi = 0$ in panel (b), respectively. In the middle column of the two panels,   
    Three-dimensional \rv{visualisations} of the square wave-pattern at $35T$ \rv{are} shown and the contour maps overlaid on the interface in the $xy$ projection \rv{in the right column}, where we evaluate the effective surface viscous tension, $\tilde \sigma_{\textit{eff}} = Bq_d (\tilde \nabla_s \cdot \tilde{\mathbf u}_s)$ in purely dilatational \rv{case ($\chi = \pi/2$)} and shear surface viscous dissipation rate, $\tilde \xi_s = 2Bq_s |\tilde {\mathbf D}_s : \tilde {\mathbf D}_s|$ in the purely \rv{shear case ($\chi = 0$).} }
    \label{fig_2}
\end{figure*}



\rv{In figures~\ref{fig_2}(a) and (b), the temporal evolution of the axial mode $\hat{\zeta}_{01}$ and the oblique mode $\hat{\zeta}_{11}$ is shown over $40$ forcing periods for $\chi=\pi/2$ (purely dilatational) and $\chi=0$ (purely shear), respectively. The growth of $\hat{\zeta}_{01}$ is similar in both limits, with the modal amplitude saturating at $\hat \zeta_{01} \approx 0.1$ for both values of $\chi$, indicating that the two-dimensional dynamics are well captured by the axial mode and are largely independent of $\chi$. In contrast, the evolution of oblique mode $\hat{\zeta}_{11}$ depends strongly on $\chi$. For $\chi=\pi/2$, $\hat{\zeta}_{11}$ exhibits rapid linear growth over approximately $15$ time periods before saturation, whereas for $\chi=0$ its growth rate is comparable to that of $\hat{\zeta}_{01}$ but saturates at a significantly lower amplitude. This behaviour indicates that surface viscosity induces three-dimensional damping effects that depend on $\chi$, with shear surface viscosity suppressing the oblique mode more effectively than dilatational surface viscosity.}

To understand the damping of the oblique mode, a three-dimensional visualisation of the interface is shown in the middle column of figure \ref{fig_2} at $t = 35T$ for $\chi =\pi/2$ and $\chi = 0$. The interface is coloured according to the magnitude of the surface velocity, $\mathbf u_s = (\mathbf I-\mathbf{nn}) \cdot \mathbf u$. The three-dimensional glyphs at the interface represent the surface velocity direction. The surface velocity indicates that the flow along the interface is from the trough (centre of the square pattern) to the crest (edges and corners of the pattern).  The surface velocity drops to zero at the troughs, corners, and middle of the edges. The radially outward $\mathbf u_s$ at the trough and inward at the middle of the edges and corners form the dilated and compressed zones of the surface, respectively. This is shown in the three-dimensional visualisation of figure \ref{fig_2}(a) for $\chi = \pi/2$. The dilated trough and compressed crest create effectively higher and lower surface viscous tensions, respectively, owing to surface compressibility. This is shown in figure \ref{fig_2}(a), where the surface is coloured by the contours of $\tilde \sigma_{\textit{eff}} = Bq_d(\tilde \nabla_s \cdot \tilde{\mathbf u}_s)$, which quantifies the additional surface tension-like stresses owing to the surface compressibility. The surface dilated zone at the centre of the square pattern showed the maximum $\tilde \sigma_{\textit{eff}}$ and minimum at the compressed corners. This results in an additional tangential stress ($\tilde \nabla_s \tilde \sigma_{\textit{eff}}$) from the crests to the troughs, which opposes the surface flow from the trough to the crest \textcolor{redstrikin}{and} is the primary reason why the threshold acceleration to destabilise a viscous surface increases with increasing $Bq$. It is interesting to note that the role of $\tilde \sigma_{\textit{eff}}$ is similar to that of surfactant-dependent surface tension, which causes the Marangoni flow from the crests to the troughs. Panda et al. \cite{panda2025marangoni} elucidated the Marangoni-driven flow mechanism in the context of square wave-patterns. Similar studies have also been conducted for spilling waves \cite{Erinin_Liu_Liu_Mostert_Deike_Duncan_2023}.   

Although the surface compressibility in the case of purely shear, $\chi = 0$, is qualitatively similar to that in the case of purely dilatational, as a square wave-pattern is formed, the surface velocity gradients differ significantly. Along the line of centre to the corners of the square wave-pattern in figure \ref{fig_2}(b), the surface velocity is evidently increased to reduce the surface velocity gradients compared to the purely dilatational case. Similar behaviour was observed by \rv{Gounley et al.} \cite{Gounley_Boedec_Jaeger_Leonetti_2016} where a neutrally buoyant drop was subjected to a constant shear rate by the background fluid. In the case of purely shear, they also observed that the surface velocity gradients were reduced, and a circular surface velocity field was observed around the drop and parallel to the direction of the shear flow. To quantify the shear surface viscous dissipation, in figure \ref{fig_2}(b), we plotted \rv{$\tilde \xi_s = Bq_s |\tilde{\mathbf D}_s : \tilde{\mathbf D}_s|$}, where it is evident that the shear surface viscous dissipation is maximum on the diagonals of the square pattern. This additional surface viscous dissipation at the diagonals is the main reason for the damping of the oblique mode, $\hat \zeta_{11}$, in the Fourier space.     

Therefore, we verified our hypothesis that three-dimensional surface rheological effects are observed in standing wave-patterns. \rv{When the pattern is transformed to Fourier space}, we can decouple the two-dimensional flows as axial modes, where the surface viscous effects are independent of $\chi$. However, the three-dimensional effects are successfully captured by the oblique modes, where additional shear surface viscous dissipation along the diagonal damp the modal amplitude of the oblique mode in the case of purely shear, compared to the purely dilatational case. In the next subsection, we extend our study to a higher-dimensional Fourier space, the six-fold hexagonal symmetry, where more than one oblique mode is found.   

 \begin{figure*}[h!]
    \centering
    \includegraphics[width=1.0\linewidth]{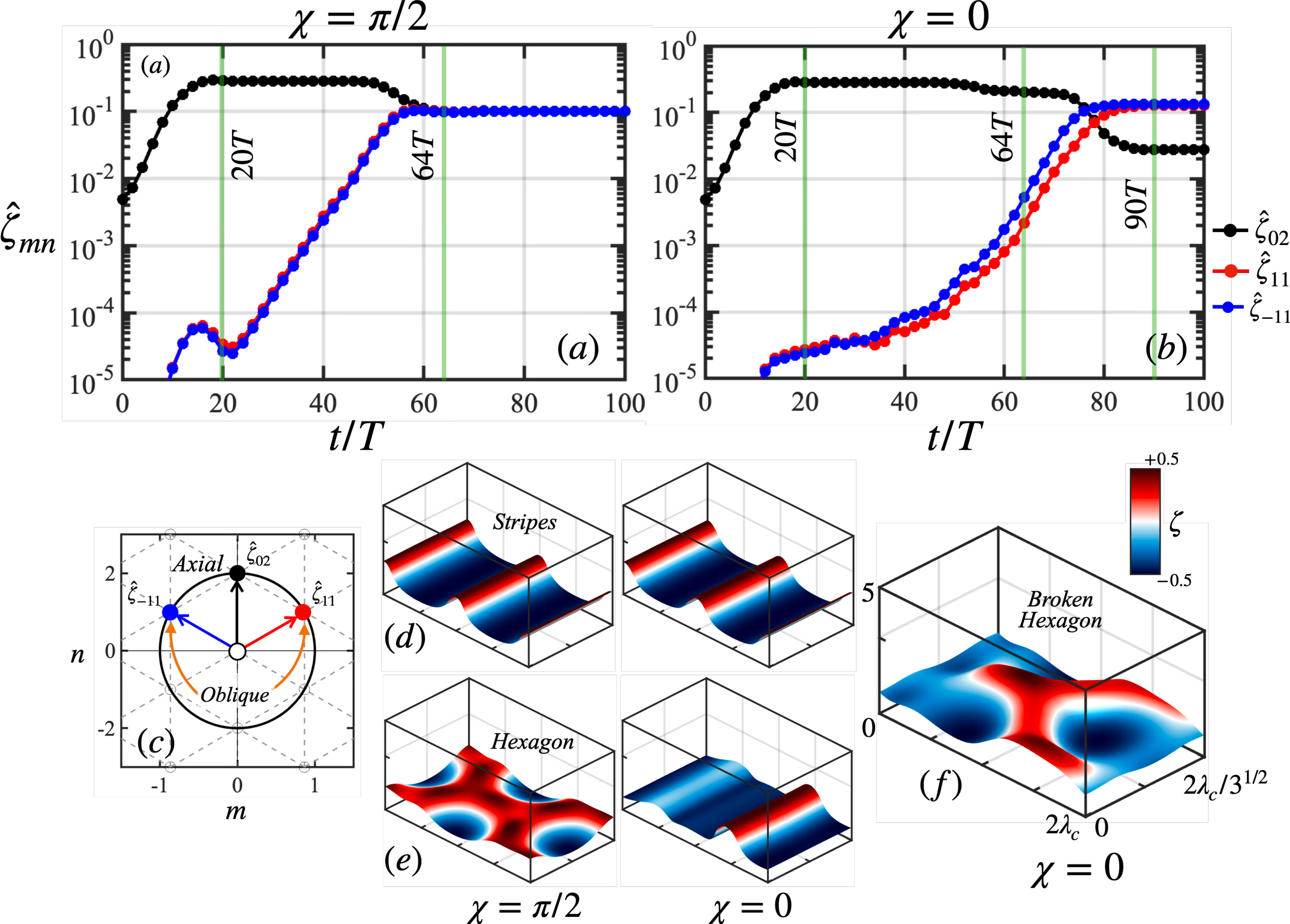}
    \caption{\rv{Evolution of non-negligible hexagonal modes in six-fold symmetry:  temporal evolution of the axial mode (black) and the oblique modes (red and blue) is shown in the interval of $100$ time periods for $Bq = 5$ and $\chi =\pi/2$ in (a) and $\chi=0$ in (b), respectively, on the basis of the Fourier space shown in (c). The green highlighted vertical lines in (a) and (b) are the time instances at which the three-dimensional visualisation of the pattern on the surface is shown in (d), (e) and (f) for $\chi = \pi/2$ (left column) and $\chi = 0$ (right column) of (d) and (e) and $\chi = 0$ for (f). Here, the interface is coloured by the interfacial vertical displacement, $\zeta$. Six-fold symmetric hexagonal pattern is only obtained in the dilatational surface viscous-    dominant case, $\chi = \pi/2$ at $t = 64T$.}}
    \label{fig_3}
\end{figure*}
 \subsection*{Secondary instabilities in hexagonal symmetry} 
 We examined the surface rheological effects in a hexagonal symmetric domain.  
 In contrast to an equilateral domain to obtain square standing wave-patterns, a hexagonal wave-pattern is obtained by employing a cuboidal computational domain whose lateral dimensions follow a \rv{$1$:$\sqrt 3$ ratio.}
 This choice enabled a six-fold compatible hexagonal Fourier space. \rv{Similar to the square wave-patterns, the temporal evolution of the non-negligible modal amplitudes are tracked in the hexagonal Fourier space as shown in figure \ref{fig_3}(a) and (b).} 
 The spatial Fourier transform is defined by,
 $\zeta(\mathbf x, t) = \sum_{m,n} e^{i\mathbf k_c \cdot \mathbf x} ~\hat \zeta_{m,n}(t)$, where a hexagon-based Fourier space is formulated as $\hat \zeta_{mn} = \sqrt 3 m~k_c/2 \mathbf e_x + n~k_c/2 \mathbf e_y$. 
 \rv{As shown in figure \ref{fig_3}(c)}, the non-negligible Fourier modes forming the six-fold symmetry are, $\hat \zeta_{02}$, \rv{an axial mode along} the $y-$ direction, 
 $\hat \zeta_{11}$, and $\hat \zeta_{-11}$, the oblique modes directing at angles of $\pi/3$ and $2\pi/3$ \rv{with respect to $\mathbf e_x$}, and their complex conjugates. 
We set \rv{$F = 22 (> F_c)$ and $Bq= 5$, }and run cases at the extrema, $\chi = 0$ and $\chi = \pi/2$, referring to the purely shear and dilatational surface viscosity cases, respectively. 
Initially, the interface was completely flat, without perturbations. 
Destabilisation is generated solely by gravity modulation (refer to the Supplemental Material). 
We begin tracking the interface when a finite amplitude was observed and set the time $t = 0$. The sampling period of the Fourier modes are set to one forcing period, such that we track the time at which the axial and oblique modal amplitudes are of the same magnitude to form a hexagon, as shown in figure \ref{fig_1}(a).   

The temporal evolution of the modal amplitudes $\hat \zeta_{02}, \hat \zeta_{11}, $ and $\hat \zeta_{-11}$ is shown in figure \ref{fig_3}(a) and (b) for the two limiting cases of surface viscosities.
Once the interface is finitely destabilised, the axial mode, $\hat \zeta_{02}$ exhibits identical growth in both cases. This demonstrates that for hexagonal wave-patterns, Fourier decomposition effectively decomposes three-dimensional patterns into two-dimensional flows, where $\mu_s$ and $\mu_d$ are indistinguishable.  
After at least $20$ forcing periods, the growth of the axial modes nonlinearly saturated, \rv{and stripe patterns are formed at both extrema of $\chi$ as shown in figure \ref{fig_3}(d) at $t = 20T$.}

The choice of waves along the axial mode is common in rectangular periodic domains, where six-fold symmetric oblique modes emerge as secondary instabilities. This is shown by the growth of $\hat \zeta_{11}$ and $\hat \zeta_{-11}$ for $t > 20T$ in figure. \ref{fig_3}(a) and (b). Although the growth of oblique modes is observed in either extrema of $\chi$, the growth is faster in the case of purely dilatational, $\chi = \pi/2$. The oblique modal amplitudes in the purely dilatational surface viscous interface grow at the same rate and nonlinearly saturate at $t \approx 60T$. The axial modal amplitude also declines, such that six-fold symmetric hexagonal waves are observed for $t \ge 60T$ as shown in figure \rv{\ref{fig_3}(e)} at $t = 64T$ for $\chi = \pi/2$. 

\rv{The oblique modal amplitudes} in the case of purely shear surface viscosity, $\chi = 0$, grow non-linearly, as also observed in the case of square wave-patterns. These modes are also found to grow at a slower rate than in the purely dilatational case. 
As the oblique modal amplitude becomes non-negligible, the amplitude of the axial mode declines, and certain irregular patterns are observed. One such instance is shown in figure \rv{\ref{fig_3}(e)} for $\chi = 0$ at $t = 64T$ where one of the crests is inhibited and the other is found to  grow. \rv{With increasing time}, a spanwise secondary instability is observed along the $x-$ direction. In Fourier space, we found that the oblique modes saturated at a higher magnitude than the axial mode (Figure \ref{fig_3}(b)) which led to the formation of broken hexagonal patterns, \rv{as shown in figure \ref{fig_3}(f)}. 

\begin{figure*}[h!]
    \centering
    \includegraphics[width=1.\linewidth]{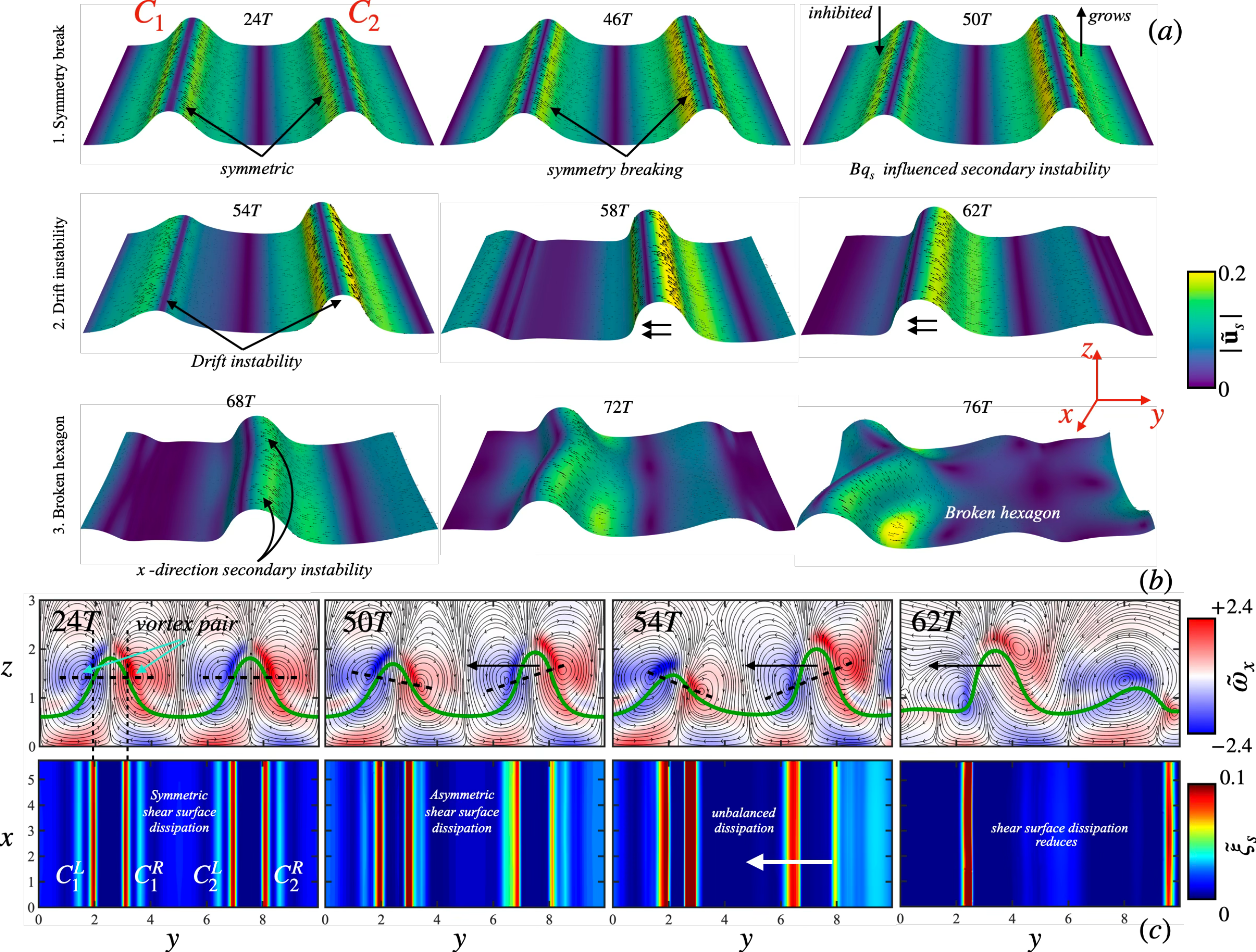}
    \caption{\rv{(a)} Three-dimensional visualisation of surface shear viscous interface coloured by the magnitude of surface velocity. The temporal evolution \rv{from stripes to broken hexagons } is categorised into \rv{three} stages: $Bq_s$-influenced secondary instability along \rv{$x-$direction stripes}, as shown in the top panel of (a); drift instability along $y-$direction, as shown in the middle; and formation of broken hexagons, as shown in the bottom panel. The $yz$ projection of the interface at $x = \lambda/\sqrt 3$ is shown in panel (b) for $t = 24T, 50T, 54T,$ and $62T$. The interface is overlaid \rv{with streamlines} and vorticity in the $x-$ direction. The black dotted line in the subfigures of panel (b) corresponds to the alignment of the vortex pair across the crests labelled $C_1$ and $C_2$. The shear surface viscous dissipation rate $\tilde \xi_s = Bq_s |\tilde{\mathbf D}_s : \tilde{\mathbf D}_s|$ is contoured on the interface in the $xy$ projection. The point of inflection of the crests $C_1$ and $C_2$ are labelled as $C_1^L$, $C_2^L$, $C_1^R$, and $C_2^R$, respectively. \rv{This corresponds to the purely shear surface viscous case of $Bq = 5$ and $\chi = 0$.} }
    \label{fig_4}
\end{figure*}

To understand the transition from regular to irregular patterns \rv{for $\chi = 0$ and $Bq = 5$}, a detailed investigation is shown in figure \ref{fig_4}(a), where the three-dimensional interface is coloured by the surface velocity and overlaid with velocity quivers for various forcing periods. In figure \ref{fig_4}(b), the $yz$ projection of the interface is shown, where the projected slice is coloured by the vorticity contours and overlaid by the velocity streamlines. The $xy$ projection of the interface is shown in figure \ref{fig_4}(c), where the interface is coloured by the shear surface viscous dissipation rate, \rv{$\tilde \xi_s =2Bq_s (\tilde{\mathbf D}_s :\tilde{\mathbf D}_s)$}.  

We designate the two crests of the stripe as $C_1$ and $C_2$ as shown in figure \ref{fig_4}(a). At $t = 24T$, $C_1$ and $C_2$ are equivalent, and symmetric regular stripes are observed. The shear surface viscous dissipation rate, $\tilde \xi_s$, is symmetric across each crest, $C_1$ and $C_2$ and is maximum at the point of inflection of the waves (figure~\ref{fig_4}(c)). These locations are designated as $C_1^L, C_1^R, C_2^L,$ and $C_2^R$, respectively. As a consequence of the equivalent $\tilde \xi_s$ at the four inflection points, the vortex pairs in figure \ref{fig_4}(b) at $t = 24T$ are equivalent in magnitude; therefore, the vortex pair across each crest is parallel to the $y-$ direction (refer the dotted line across the crests in Figure \ref{fig_4}(c)). At $t = 50T$, $\tilde \xi_s$ is unequal in the inflection zones, as shown in figure \ref{fig_4}(c). This is because the shear surface viscous dissipation is maximum at $C_1$ compared to that at $C_2$. This leads to the inhibition of $C_1$ and the growth of $C_2$ as shown in figure \ref{fig_4}(a). Thus, we observe a \rv{slight} dip in the modal amplitude of $\hat \zeta_{02}$ in figure \ref{fig_3}(b) for $t \approx 50T$. 

The shear surface viscous dissipation is not only asymmetric to the crests but also across each crest. For instance, $\tilde \xi_s$ is higher at $C_1^R$ than $C_1^L$. This leads to the inclination of the vortex pair across $C_1$ as shown in figure \ref{fig_4}(b) at $t = 50T$. 
Because $\tilde \xi_s$ across $C_2$ is lower than $C_1$ and the inclination of the vortex pairs, a horizontally directed motion is initiated, whereas the external vibration is solely vertical. This phenomenon is known as \rv{`drift instability’, which } has been previously reported in narrow channels \cite{kucher2025discovery}, mixed forcing schemes \cite{marin2021drift}, odd viscosity \cite{chu2022effect}, and long-time wall shear effects \cite{martin2002drift}. Drift instability has also been theoretically acknowledged in the case of a viscous surface interface \cite{martin2006effect}; however, in this study, we present, for the first time, the mechanism of drift instability due to asymmetric $\tilde \xi_s$ using direct numerical simulation. 

As time progresses, the dissipation increases on $C_1$, as compared to $C_2$ (see $\tilde \xi_s$ at $t=54T$ and $62T$ in figure \ref{fig_4}(b) and (c)), \rv{respectively, until a} spanwise secondary instability along the $x-$ direction is observed at $t=68T$ (see figure. \ref{fig_4}(a)), \rv{which} is a consequence of the growing oblique modes, as shown in figure \ref{fig_3}(b). However, unlike the balanced saturation of $\hat \zeta_{02}, \hat \zeta_{11}$ and $\hat \zeta_{-11}$ to form the hexagonal wave-pattern in $\chi = 0$, 
\rv{the} growth of the oblique modes, $\hat \zeta_{-11}$ and $\hat \zeta_{11}$ is fed by the inhibition of the $\zeta_{02}$ in the case of $\chi = \pi/2$. This leads to a broken hexagonal pattern, where the modal amplitude of the oblique modes is different from that of the axial mode. The difference between the axial and oblique modes is due to the additional shear surface viscous dissipation along the $x-$ direction, which stabilises the waves against oblique modes for a longer period of time. This is why the axial and oblique modes compete with each other. 

\begin{figure}[h!]
    \centering \includegraphics[width=0.65\linewidth]{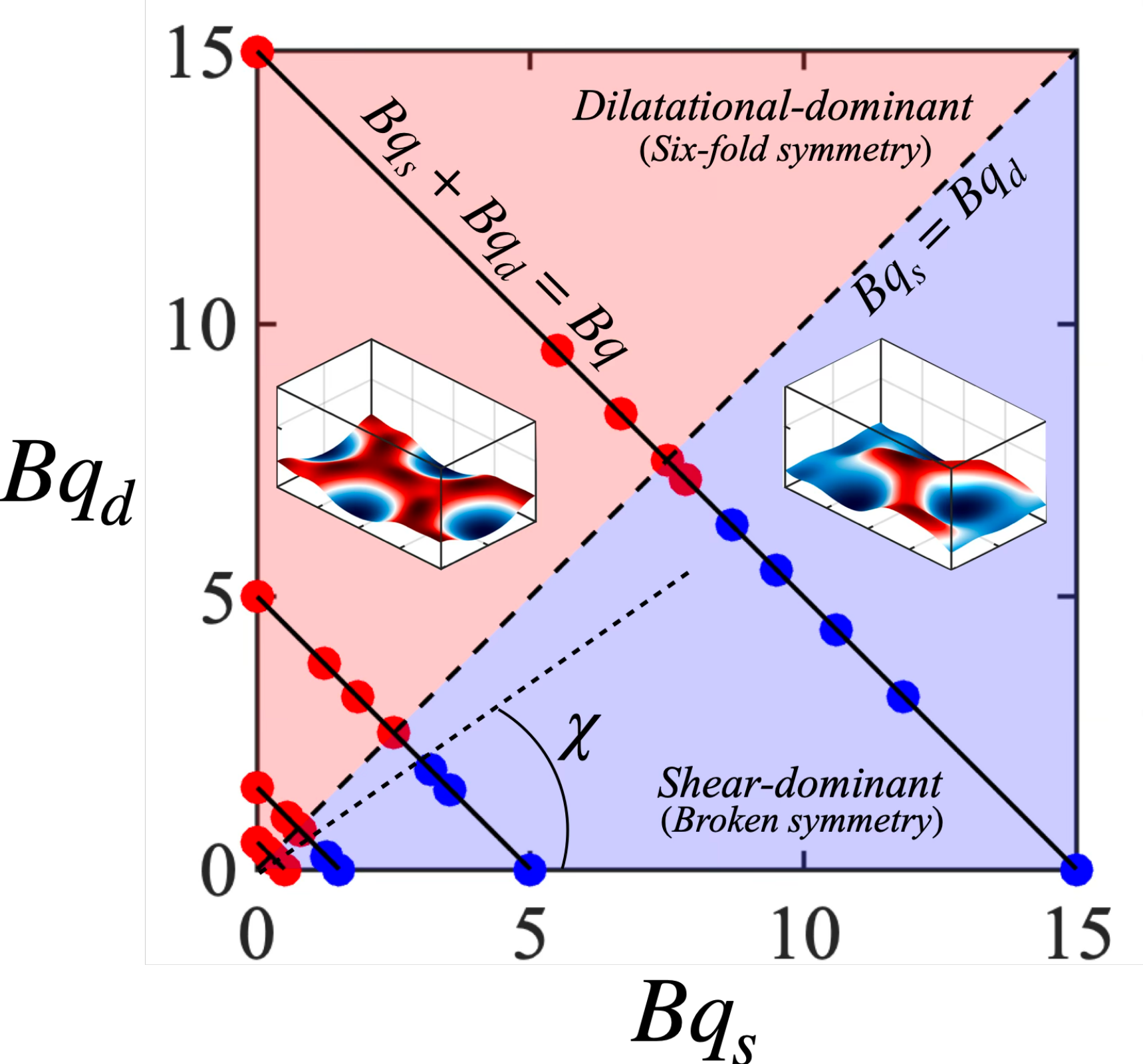}
    \caption{Regime map: \rv{transition} from a six-fold symmetric pattern to a non-symmetric broken hexagonal structures. Here, the blue region ($\chi < \pi/4$), filled with blue circles corresponds to the shear surface \rv{viscous-dominated} interface and the red ($\chi \ge \pi/4$) region filled with red circles refers to the dilatational surface \rv{viscous-dominated} interface. \rv{Three-dimensional visualisations of the hexagon and broken-hexagon are shown in the inset.}}
    \label{fig_5}
\end{figure}

In figure \ref{fig_5}, several numerical experiments are carried out for $Bq= 
[0.5, 15]$ and $\chi = [0, \pi/2]$ \rv{and the outcomes are summarised in the $(Bq_s,Bq_d)$ space}.  For lower $Bq$, a hexagonal pattern is observed for all $\chi$. However, as we increase $Bq$, broken hexagons are obtained for lower $\chi$, that is, shear surface viscous dominant regime. As we increase $Bq$, the pattern transition from a hexagonal to a non-hexagonal pattern is observed for $\chi \approx \pi/4$. This implies that the broken hexagonal wave-pattern is a consequence of shear surface viscous dissipation along the inflection zones of the crests which inhibits the growth of the spanwise instability along the $x-$ axis. This leads to a competing modal state in the six-fold symmetry, where the oblique modes are fed by the axial mode. The limitation of obtaining regular patterns in the case of a purely shear surface viscous interface is an indication that can be exploited as a fast screening method to determine the dominant factor of surface viscosity.   

\section*{Conclusion}
In this study, we investigated the three-dimensional surface rheological effects in the context of a pattern-forming interface owing to vertical vibration. The dilatational and shear surface viscosities are highly correlated and indistinguishable when the flow is either two-dimensional or linear. By introducing two control parameters, the total Boussinesq number, $Bq = Bq_s + Bq_d$ and the mixing angle, $\tan \chi = Bq_d/Bq_s$, we first studied the effects of surface rheological effects in the linear and two-dimensional regimes. For varying $\chi$, symmetry (two-fold quasi-2D, four-fold square, six-fold hexagon), we found that the threshold acceleration $F_c$ and the linear growth rate were independent of the geometry as well as shear and dilatational surface viscosity. Second, by considering an equilateral domain, we \rv{examined} the square wave-pattern for the two limits, $\chi = 0$ and $\pi/2$. By decomposing the pattern in the spatial Fourier transform, we found that the axial modes successfully decomposed the three-dimensional flows into two-dimensional flows, where the amplitude of the axial mode was indistinguishable in the two limits. However, we found that the axial mode amplitude in the case of purely shear surface viscosity is lower than that in the case of purely dilatational viscosity due to additional shear surface viscous dissipation along the inflection zone of the surface waves. To further verify our observation, we considered a six-fold symmetry, where there were two oblique modes. Additional shear surface viscous dissipation at the inflection zone occurring along the $x-$ direction opposes spanwise instability owing to the growth of the oblique modes. Mode competition leads to the feeding of oblique modes by the axial mode, resulting in a broken hexagon pattern owing to the dominant shear surface viscosity. Finally, we conducted a parametric study of varying $Bq$ and $\chi$, which concluded that the dominance of shear surface viscosity results in a pattern transition from hexagonal wave-patterns to broken/irregular waves. This demonstrates that pattern-forming systems can be exploited as a fast screening method to understand the effects of shear and dilatational surface viscosities.  

\noindent
\section*{Methods}
\subsection*{Governing equations}
We consider a three-dimensional cuboidal computational box; laterally periodic and closed from top and bottom sides, with fixed height $H = 5~\rm mm$. To compare the dimensional effects, the lateral extent was chosen as $\lambda_c \times \lambda_c/4$, $\lambda_c \times \lambda_c$, and $2\lambda_c \times 2\lambda_c/\sqrt 3$, \rv{where $\lambda_c$ is the critical wavelength of the parametric surface waves}. The domain is filled up to $h=1~\rm{mm}$ with a water–glycerine mixture ($\rho_l=1000~\rm{kg/m}^3$, $\mu_l=0.025 ~\text{Pa. s}$), initially flat and at rest, with air above ($\rho_a=1.205~\rm{kg/m}^3$, $\mu_a=1.82\times10^{-5}~\text{Pa. s}$). The liquid–air interface with surface tension $\sigma$ and Newtonian surface rheology characterised by shear viscosity $\mu_s$ and dilatational viscosity $\mu_d$ is subjected to vibration in the positive $z-$ direction. In the interface frame, the vibrations appear as a parametric modulation of the volumetric forces with amplitude A and frequency $\Omega$. Non-dimensionalisation with length scaled by \rv{$h$}, time by $1/\Omega$, density by $\rho_l$, and viscosity by $\mu_l$ yields the one-fluid incompressible two-phase formulation, given by,    
\begin{subequations}
\begin{equation}
    \tilde \nabla \cdot \tilde {\mathbf u} = 0,
    \label{divu}
\end{equation}
\begin{align}
    \tilde \rho \dfrac{{\rm D} \tilde{\mathbf u}}{{\rm D}\tilde t} &= \tilde {\mathbf \nabla} \cdot \left(-\tilde p \mathbf I ~+~ \frac{\tilde \mu }{Re} \left(\tilde \nabla \tilde{\mathbf u} + (\tilde \nabla \tilde{\mathbf u})^{\rm T}\right)\right)  + \frac{\tilde \rho}{Fr^2}\left(-1 + F\cos \tilde t\right)\mathbf e_z  \nonumber \\
    & + \int_{\tilde A'} \tilde \nabla_s \cdot \left[\dfrac{\mathbf I_s}{We} + \dfrac{Bq_d-Bq_s}{Re}(\tilde \nabla_s \cdot \tilde {\mathbf u}) \mathbf I_s + \frac{Bq_s}{Re}  (\tilde \nabla_s \tilde{\mathbf u} \cdot \mathbf I_s ~+~ \mathbf I_s \cdot (\tilde \nabla_s \tilde {\mathbf u})^{\mathrm T}) \right]  
    ~\tilde \delta(\tilde {\mathbf x}- \tilde{\mathbf x}_f) d\tilde A'.
    \label{full-nse}
\end{align}
\end{subequations}
Here, $\tilde \rho$ and $\tilde \mu$ are one-fluid density material properties defined by the Heaviside function $\mathcal H(\tilde{\mathbf x}, \tilde t)$, which is $1$ for water and $0$ for air. The one-fluid material properties are given by, 
\begin{subequations}
    \begin{equation}
    \tilde \rho = \mathcal H(\tilde {\mathbf x}, \tilde t) + (\rho_a/\rho_l) (1 - \mathcal H(\tilde {\mathbf x}, \tilde t))
    \end{equation}
    \begin{equation}
    \tilde \mu = \mathcal H(\tilde {\mathbf x}, \tilde t) + (\mu_a/\mu_l) (1 - \mathcal H(\tilde {\mathbf x}, \tilde t))
    \end{equation}
\end{subequations}
$\tilde \delta(\tilde{\mathbf x} - \tilde{\mathbf x}_f)$ is the \rv{Dirac} delta function, which is non-zero if and only if $\mathbf x = \mathbf x_f$; $\nabla_s (\nabla_s \cdot)$ is the surface gradient (divergence) operator and $\mathbf I_s = \mathbf I - \mathbf n\mathbf n$ is the surface identity tensor, where $\mathbf n$ is the unit normal vector to the interface. The dimensionless numbers arising in the governing equations are Reynolds ($Re$), driving amplitude ($F$), Froude ($Fr$), Weber ($We$), dilatational and shear Boussinesq ($Bq_s$ and $Bq_d$) numbers, given by,    
\begin{equation}
    Re = \frac{\rho_l \Omega^2h^3}{\mu_l}, ~~ Fr = \Omega \sqrt\frac{h}{g}, ~~ F = \frac{A}{g}, ~~ We = \frac{\rho_l \Omega^2 h^3}{\sigma}, ~~ Bq_d = \frac{\mu_d }{\mu_l h}, ~~ Bq_s =\frac{\mu_s}{\mu_l h}.   
\end{equation}
\rv{For a system dominated by competing vibratory inertial, capillary, and viscous forces,} we set $Re \approx 25, ~ We \approx 7,$ and $Fr \approx 6$ by setting the angular frequency $\Omega = 200\pi ~\rm{rad~s^{-1}}$. 

\rv{\subsection*{Numerical method}}
\noindent Equations \eqref{divu} and \eqref{full-nse} are solved using a hybrid level-set-based front-tracking method known as the level contour reconstruction method (LCRM), which is well documented in the literature \cite{shin2002modeling, shin2018jcp, shin2017solver}.
The spatial derivatives on the Eulerian grid were calculated using a standard cell-centred scheme, except for the nonlinear convective term, for which we implemented an essentially non-oscillatory (ENO) procedure on a staggered grid. Peskin's immersed boundary method was used to couple the Eulerian and Lagrangian grids.
The advection of the Lagrangian field $\mathbf x_f(t+\Delta t) = \int_t^{t+\Delta t} \mathbf u_f(t) dt$, where $\mathbf u_f(t)$ is the interpolated velocity at the interface at time $t$, is accomplished by second-order Runge-Kutta numerical integration. A resolution of $|\Delta x| = |\Delta y| = \lambda_c/44$ was found to be necessary to capture the Faraday wave dynamics in \cite{perinet2009numerical, kahouadji2015numerical}. 
We chose a finer resolution of $\lambda_c/128$ to capture the coupling with the surface rheological effects. \rv{The time-stepping was fixed at $\Delta t = 10^{-8} ~ \rm s$ for $Bq \ge 1$ and $10^{-7}~ \rm s$ for $Bq < 1$.}

After benchmarking surfactant transport on a highly deforming interface \cite{shin2018jcp}, the numerical model was extended and validated for surface rheological effects \cite{panda2025directnumericalsimulationtwophase}. In this article, we briefly describe the implementation of surface rheology. 
\rv{The surface dilatational viscous contribution is reformulated in terms of a surface-viscous tension form, given by, $\sigma_{vis} = \sigma + (\mu_d - \mu_s)(\nabla_s \cdot \mathbf u)$, enabling an evaluation analogous to the classical surface tension forces. This yields a normal component proportional to the local curvature and a tangential component owing to the surface gradients of $\sigma_{vis}$. Whereas the normal components are directly computed on the Lagrangian elements, the surface gradient of $\sigma_{vis}$ is evaluated using a probing technique by interpolation on either side of the Lagrangian element. Both contributions are transferred to the Eulerian grid by Peskin's immersed boundary method \cite{peskin1995general}.}

\rv{Shear surface viscous forces are evaluated using the extrinsic-Lagrangian formulation of the surface rate-of-deformation tensor, obtained by projecting the interpolated bulk velocity gradient onto the local tangent plane of the interface. The resulting surface deformation tensor is integrated along the edges of each interfacial element to compute the shear viscous force. This approach ensures a consistent and robust treatment of surface momentum diffusion and allows for the accurate coupling of interfacial rheology with bulk hydrodynamics under large deformations and topological changes in the interface. }

To accurately calculate the surface viscous effects, the main objective is to evaluate tensor $\nabla_s \mathbf u$. Leveraging the capabilities of LCRM, the $\nabla \mathbf u$ tensor is readily evaluated on the Eulerian grid and then interpolated to the centre of the triangular faces of the Lagrangian grid. Using the information of $\mathbf n$ obtained from the level set function at the centre of the triangular grid, we evaluate $\nabla_s \mathbf u = \nabla \mathbf u - \mathbf n (\mathbf n \cdot \nabla \mathbf u)$. The surface divergence of $\mathbf u$ is evaluated by the trace of the $\nabla_s \mathbf u$ tensor, given by $\nabla_s \cdot \mathbf u = \rm{tr}(\nabla_s \mathbf u)$. The calculation of dilatational and shear surface viscous forces is straightforward on the Lagrangian interface, as discussed in detail in reference \cite{panda2025directnumericalsimulationtwophase}. 
\\
\section*{Data availability}
All data supporting the findings of this study are available from the corresponding author upon reasonable requests.
\section*{Acknowledgements}This work was supported by the Engineering and Physical Sciences Research Council, UK, through the  PREMIERE (EP/T000414/1) programme grant, the ANTENNA Prosperity Partnership grant (EP/V056891/1), and by the National Research Foundation of Korea(NRF) grant funded by the Korea government (MSIT) (No. RS-2025-02302984). We acknowledge the HPC facilities provided by the Imperial College London Research Computing Service. We acknowledge Dr. Paula Pico for her efforts in introducing surface viscosity during her doctoral studies. D.P. acknowledges the Imperial College London President’s PhD Scholarship. A.M.A. acknowledges the Kuwait Foundation for the Advancement of Sciences (KFAS) for its financial support. D.J. and J.C. acknowledge support through HPC/AI computing time at the Institut du Developpement et des Ressources en Informatique Scientifique (IDRIS) of the Centre National de la Recherche Scientifique (CNRS), coordinated by GENCI (Grand Equipement National de Calcul Intensif) 
grant 2025 A0182B06721. The numerical simulations were performed using the BLUE code \cite{shin2017solver} and the visualisations were generated using ParaView. 

\section*{Author contributions}
D.P. was involved in conceptualisation, performing simulations, and writing the initial draft. A.M.A, M.S, and L.K., were involved in analysing, data acquisition, writing and editing. J.C, S.S, D.J were involved in supervision, editing, and review. O.K.M supervised, acquired funding, and was involved in conceptualisation, editing and review.

\begin{thebibliography}{10}
\expandafter\ifx\csname url\endcsname\relax
  \def\url#1{\texttt{#1}}\fi
\expandafter\ifx\csname urlprefix\endcsname\relax\def\urlprefix{URL }\fi
\providecommand{\bibinfo}[2]{#2}
\providecommand{\eprint}[2][]{\url{#2}}

\bibitem{fuller2012complex}
\bibinfo{author}{Fuller, G.~G.} \& \bibinfo{author}{Vermant, J.}
\newblock \bibinfo{title}{Complex fluid-fluid interfaces: rheology and structure}.
\newblock \emph{\bibinfo{journal}{Annual review of chemical and biomolecular engineering}} \textbf{\bibinfo{volume}{3}}, \bibinfo{pages}{519--543} (\bibinfo{year}{2012}).

\bibitem{miller2010rheology}
\bibinfo{author}{Miller, R.} \emph{et~al.}
\newblock \bibinfo{title}{Rheology of interfacial layers}.
\newblock \emph{\bibinfo{journal}{Colloid and Polymer Science}} \textbf{\bibinfo{volume}{288}}, \bibinfo{pages}{937--950} (\bibinfo{year}{2010}).

\bibitem{choi2011active}
\bibinfo{author}{Choi, S.}, \bibinfo{author}{Steltenkamp, S.}, \bibinfo{author}{Zasadzinski, J.} \& \bibinfo{author}{Squires, T.}
\newblock \bibinfo{title}{Active microrheology and simultaneous visualization of sheared phospholipid monolayers}.
\newblock \emph{\bibinfo{journal}{Nature communications}} \textbf{\bibinfo{volume}{2}}, \bibinfo{pages}{312} (\bibinfo{year}{2011}).

\bibitem{jaensson2021computational}
\bibinfo{author}{Jaensson, N.~O.}, \bibinfo{author}{Anderson, P.~D.} \& \bibinfo{author}{Vermant, J.}
\newblock \bibinfo{title}{Computational interfacial rheology}.
\newblock \emph{\bibinfo{journal}{Journal of Non-Newtonian Fluid Mechanics}} \textbf{\bibinfo{volume}{290}}, \bibinfo{pages}{104507} (\bibinfo{year}{2021}).

\bibitem{ji2023secondary}
\bibinfo{author}{Ji, B.}, \bibinfo{author}{Yang, Z.}, \bibinfo{author}{Wang, Z.}, \bibinfo{author}{Ewoldt, R.~H.} \& \bibinfo{author}{Feng, J.}
\newblock \bibinfo{title}{Secondary bubble entrainment via primary bubble bursting at a viscoelastic surface}.
\newblock \emph{\bibinfo{journal}{Physical review letters}} \textbf{\bibinfo{volume}{131}}, \bibinfo{pages}{104002} (\bibinfo{year}{2023}).

\bibitem{tammaro2021flowering}
\bibinfo{author}{Tammaro, D.} \emph{et~al.}
\newblock \bibinfo{title}{Flowering in bursting bubbles with viscoelastic interfaces}.
\newblock \emph{\bibinfo{journal}{Proceedings of the National Academy of Sciences}} \textbf{\bibinfo{volume}{118}}, \bibinfo{pages}{e2105058118} (\bibinfo{year}{2021}).

\bibitem{erni2011deformation}
\bibinfo{author}{Erni, P.}
\newblock \bibinfo{title}{Deformation modes of complex fluid interfaces}.
\newblock \emph{\bibinfo{journal}{Soft Matter}} \textbf{\bibinfo{volume}{7}}, \bibinfo{pages}{7586--7600} (\bibinfo{year}{2011}).

\bibitem{lopez1998direct}
\bibinfo{author}{Lopez, J.} \& \bibinfo{author}{Hirsa, A.}
\newblock \bibinfo{title}{Direct determination of the dependence of the surface shear and dilatational viscosities on the thermodynamic state of the interface: Theoretical foundations}.
\newblock \emph{\bibinfo{journal}{Journal of colloid and interface science}} \textbf{\bibinfo{volume}{206}}, \bibinfo{pages}{231--239} (\bibinfo{year}{1998}).

\bibitem{flumerfelt1980effects}
\bibinfo{author}{Flumerfelt, R.~W.}
\newblock \bibinfo{title}{Effects of dynamic interfacial properties on drop deformation and orientation in shear and extensional flow fields}.
\newblock \emph{\bibinfo{journal}{Journal of colloid and interface science}} \textbf{\bibinfo{volume}{76}}, \bibinfo{pages}{330--349} (\bibinfo{year}{1980}).

\bibitem{singh2020deformation}
\bibinfo{author}{Singh, N.} \& \bibinfo{author}{Narsimhan, V.}
\newblock \bibinfo{title}{Deformation and burst of a liquid droplet with viscous surface moduli in a linear flow field}.
\newblock \emph{\bibinfo{journal}{Physical Review Fluids}} \textbf{\bibinfo{volume}{5}}, \bibinfo{pages}{063601} (\bibinfo{year}{2020}).

\bibitem{ubal2005influence}
\bibinfo{author}{Ubal, S.}, \bibinfo{author}{Giavedoni, M.~D.} \& \bibinfo{author}{Saita, F.~A.}
\newblock \bibinfo{title}{Influence of surface viscosity on two-dimensional faraday waves}.
\newblock \emph{\bibinfo{journal}{Industrial \& engineering chemistry research}} \textbf{\bibinfo{volume}{44}}, \bibinfo{pages}{1090--1099} (\bibinfo{year}{2005}).

\bibitem{wee2022absence}
\bibinfo{author}{Wee, H.}, \bibinfo{author}{Wagoner, B.~W.} \& \bibinfo{author}{Basaran, O.~A.}
\newblock \bibinfo{title}{Absence of scaling transitions in breakup of liquid jets caused by surface viscosity}.
\newblock \emph{\bibinfo{journal}{Physical Review Fluids}} \textbf{\bibinfo{volume}{7}}, \bibinfo{pages}{074001} (\bibinfo{year}{2022}).

\bibitem{perinet2009numerical}
\bibinfo{author}{P{\'e}rinet, N.}, \bibinfo{author}{Juric, D.} \& \bibinfo{author}{Tuckerman, L.}
\newblock \bibinfo{title}{Numerical simulation of {Faraday} waves}.
\newblock \emph{\bibinfo{journal}{J.\ Fluid Mech.}} \textbf{\bibinfo{volume}{635}}, \bibinfo{pages}{1--26} (\bibinfo{year}{2009}).

\bibitem{edwards1993parametrically}
\bibinfo{author}{Edwards, W.} \& \bibinfo{author}{Fauve, S.}
\newblock \bibinfo{title}{Parametrically excited quasicrystalline surface waves}.
\newblock \emph{\bibinfo{journal}{Phys.\ Rev.\ E}} \textbf{\bibinfo{volume}{47}}, \bibinfo{pages}{R788} (\bibinfo{year}{1993}).

\bibitem{kudrolli1998superlattice}
\bibinfo{author}{Kudrolli, A.}, \bibinfo{author}{Pier, B.} \& \bibinfo{author}{Gollub, J.}
\newblock \bibinfo{title}{Superlattice patterns in surface waves}.
\newblock \emph{\bibinfo{journal}{Physica D}} \textbf{\bibinfo{volume}{123}}, \bibinfo{pages}{99--111} (\bibinfo{year}{1998}).

\bibitem{henderson1998effects}
\bibinfo{author}{Henderson, D.~M.}
\newblock \bibinfo{title}{Effects of surfactants on faraday-wave dynamics}.
\newblock \emph{\bibinfo{journal}{Journal of Fluid Mechanics}} \textbf{\bibinfo{volume}{365}}, \bibinfo{pages}{89--107} (\bibinfo{year}{1998}).

\bibitem{merkt2004persistent}
\bibinfo{author}{Merkt, F.~S.}, \bibinfo{author}{Deegan, R.~D.}, \bibinfo{author}{Goldman, D.~I.}, \bibinfo{author}{Rericha, E.~C.} \& \bibinfo{author}{Swinney, H.~L.}
\newblock \bibinfo{title}{Persistent holes in a fluid}.
\newblock \emph{\bibinfo{journal}{Physical review letters}} \textbf{\bibinfo{volume}{92}}, \bibinfo{pages}{184501} (\bibinfo{year}{2004}).

\bibitem{panda2025marangoni}
\bibinfo{author}{Panda, D.} \emph{et~al.}
\newblock \bibinfo{title}{Marangoni-driven patterns, ridges and hills in surfactant-covered parametric surface waves}.
\newblock \emph{\bibinfo{journal}{Journal of Fluid Mechanics}} \textbf{\bibinfo{volume}{1008}}, \bibinfo{pages}{R4} (\bibinfo{year}{2025}).

\bibitem{scriven1960dynamics}
\bibinfo{author}{Scriven, L.~E.}
\newblock \bibinfo{title}{Dynamics of a fluid interface equation of motion for newtonian surface fluids}.
\newblock \emph{\bibinfo{journal}{Chemical Engineering Science}} \textbf{\bibinfo{volume}{12}}, \bibinfo{pages}{98--108} (\bibinfo{year}{1960}).

\bibitem{Gounley_Boedec_Jaeger_Leonetti_2016}
\bibinfo{author}{Gounley, J.}, \bibinfo{author}{Boedec, G.}, \bibinfo{author}{Jaeger, M.} \& \bibinfo{author}{Leonetti, M.}
\newblock \bibinfo{title}{Influence of surface viscosity on droplets in shear flow}.
\newblock \emph{\bibinfo{journal}{Journal of Fluid Mechanics}} \textbf{\bibinfo{volume}{791}}, \bibinfo{pages}{464–494} (\bibinfo{year}{2016}).

\bibitem{erni2011emulsion}
\bibinfo{author}{Erni, P.}, \bibinfo{author}{Windhab, E.~J.} \& \bibinfo{author}{Fischer, P.}
\newblock \bibinfo{title}{Emulsion drops with complex interfaces: Globular versus flexible proteins}.
\newblock \emph{\bibinfo{journal}{Macromolecular materials and engineering}} \textbf{\bibinfo{volume}{296}}, \bibinfo{pages}{249--262} (\bibinfo{year}{2011}).

\bibitem{kragel1996surface}
\bibinfo{author}{Kr{\"a}gel, J.} \emph{et~al.}
\newblock \bibinfo{title}{Surface rheology of monolayers}.
\newblock \emph{\bibinfo{journal}{Thin Solid Films}} \textbf{\bibinfo{volume}{284}}, \bibinfo{pages}{361--364} (\bibinfo{year}{1996}).

\bibitem{brenner2013interfacial}
\bibinfo{author}{Brenner, H.}
\newblock \emph{\bibinfo{title}{Interfacial transport processes and rheology}} (\bibinfo{publisher}{Elsevier}, \bibinfo{year}{2013}).

\bibitem{ubal2005SV}
\bibinfo{author}{Ubal, S.}, \bibinfo{author}{Giavedoni, M.} \& \bibinfo{author}{Saita, F.}
\newblock \bibinfo{title}{Influence of surface viscosity on two-dimensional {Faraday} waves}.
\newblock \emph{\bibinfo{journal}{Industrial \& engineering chemistry research}} \textbf{\bibinfo{volume}{44}}, \bibinfo{pages}{1090--1099} (\bibinfo{year}{2005}).

\bibitem{Erinin_Liu_Liu_Mostert_Deike_Duncan_2023}
\bibinfo{author}{Erinin, M.} \emph{et~al.}
\newblock \bibinfo{title}{The effects of surfactants on plunging breakers}.
\newblock \emph{\bibinfo{journal}{J.\ Fluid Mech.}} \textbf{\bibinfo{volume}{972}}, \bibinfo{pages}{R5} (\bibinfo{year}{2023}).

\bibitem{kucher2025discovery}
\bibinfo{author}{Kucher, S.}, \bibinfo{author}{Wesfreid, J.~E.} \& \bibinfo{author}{Cobelli, P.~J.}
\newblock \bibinfo{title}{Discovery of propagating trains of oscillons over faraday waves in a 1d experiment}.
\newblock \emph{\bibinfo{journal}{Europhysics Letters}} \textbf{\bibinfo{volume}{150}}, \bibinfo{pages}{33003} (\bibinfo{year}{2025}).

\bibitem{marin2021drift}
\bibinfo{author}{Mar{\'\i}n, J.~F.}, \bibinfo{author}{{\'A}vila, R.~R.}, \bibinfo{author}{Coulibaly, S.}, \bibinfo{author}{Taki, M.} \& \bibinfo{author}{Garc{\'\i}a-{\~N}ustes, M.~A.}
\newblock \bibinfo{title}{Drift instabilities in localised faraday patterns}.
\newblock \emph{\bibinfo{journal}{arXiv preprint arXiv:2112.03866}}  (\bibinfo{year}{2021}).

\bibitem{chu2022effect}
\bibinfo{author}{Chu, X.}, \bibinfo{author}{Chang, L.}, \bibinfo{author}{Jia, B.} \& \bibinfo{author}{Jian, Y.}
\newblock \bibinfo{title}{Effect of the odd viscosity on {Faraday} wave instability}.
\newblock \emph{\bibinfo{journal}{Phys.\ Fluids}} \textbf{\bibinfo{volume}{34}} (\bibinfo{year}{2022}).

\bibitem{martin2002drift}
\bibinfo{author}{Mart{\'\i}n, E.}, \bibinfo{author}{Martel, C.} \& \bibinfo{author}{Vega, J.~M.}
\newblock \bibinfo{title}{Drift instability of standing faraday waves}.
\newblock \emph{\bibinfo{journal}{Journal of Fluid Mechanics}} \textbf{\bibinfo{volume}{467}}, \bibinfo{pages}{57--79} (\bibinfo{year}{2002}).

\bibitem{martin2006effect}
\bibinfo{author}{Mart{\'\i}n, E.} \& \bibinfo{author}{Vega, J.~M.}
\newblock \bibinfo{title}{The effect of surface contamination on the drift instability of standing faraday waves}.
\newblock \emph{\bibinfo{journal}{Journal of Fluid Mechanics}} \textbf{\bibinfo{volume}{546}}, \bibinfo{pages}{203--225} (\bibinfo{year}{2006}).

\bibitem{shin2002modeling}
\bibinfo{author}{Shin, S.} \& \bibinfo{author}{Juric, D.}
\newblock \bibinfo{title}{Modeling three-dimensional multiphase flow using a level contour reconstruction method for front tracking without connectivity}.
\newblock \emph{\bibinfo{journal}{J.\ Comput.\ Phys.}} \textbf{\bibinfo{volume}{180}}, \bibinfo{pages}{427--470} (\bibinfo{year}{2002}).

\bibitem{shin2018jcp}
\bibinfo{author}{Shin, S.} \emph{et~al.}
\newblock \bibinfo{title}{A hybrid interface tracking – level set technique for multiphase flow with soluble surfactant}.
\newblock \emph{\bibinfo{journal}{J.\ Comput\ Phys.}} \textbf{\bibinfo{volume}{359}}, \bibinfo{pages}{409–435} (\bibinfo{year}{2018}).

\bibitem{shin2017solver}
\bibinfo{author}{Shin, S.}, \bibinfo{author}{Chergui, J.} \& \bibinfo{author}{Juric, D.}
\newblock \bibinfo{title}{A solver for massively parallel direct numerical simulation of three-dimensional multiphase flows}.
\newblock \emph{\bibinfo{journal}{J.\ Mech.\ Sci.\ Technol.}} \textbf{\bibinfo{volume}{31}}, \bibinfo{pages}{1739--1751} (\bibinfo{year}{2017}).

\bibitem{kahouadji2015numerical}
\bibinfo{author}{Kahouadji, L.} \emph{et~al.}
\newblock \bibinfo{title}{Numerical simulation of supersquare patterns in {Faraday} waves}.
\newblock \emph{\bibinfo{journal}{J.\ Fluid Mech.}} \textbf{\bibinfo{volume}{772}}, \bibinfo{pages}{R2} (\bibinfo{year}{2015}).

\bibitem{panda2025directnumericalsimulationtwophase}
\bibinfo{author}{Panda, D.} \emph{et~al.}
\newblock \bibinfo{title}{Direct numerical simulation of two-phase flows with surfactant-induced surface viscous effects} (\bibinfo{year}{2025}).
\newblock \urlprefix\url{https://arxiv.org/abs/2509.24722}.
\newblock \eprint{2509.24722}.

\bibitem{peskin1995general}
\bibinfo{author}{Peskin, C.~S.} \& \bibinfo{author}{McQueen, D.~M.}
\newblock \bibinfo{title}{A general method for the computer simulation of biological systems interacting with fluids.}
\newblock In \emph{\bibinfo{booktitle}{Symposia of the society for Experimental Biology}}, vol.~\bibinfo{volume}{49}, \bibinfo{pages}{265--276} (\bibinfo{year}{1995}).

\end{thebibliography}

\end{document}